\documentclass[aps,prl,twocolumn,superscriptaddress]{revtex4-2}
\usepackage{graphicx}
\usepackage{amsmath,amssymb,wasysym}
\usepackage{dsfont} 
\usepackage{color}
\usepackage[linktocpage=true,colorlinks=true, pdfborder={0 0 0},linkcolor=blue,citecolor=blue,urlcolor=blue,anchorcolor=black,linkcolor=blue]{hyperref}

\begin{document}

\title{Exceptional Points and Lasing Thresholds: When Lower-Q Modes Win} 

\author{Julius Kullig}
\affiliation{Institut f{\"u}r Physik,
  Otto-von-Guericke-Universit{\"a}t Magdeburg, Postfach 4120, D-39016
  Magdeburg, Germany}

\author{Qi Zhong}
\affiliation{Department of Electrical and Computer Engineering, Saint Louis University, Saint Louis, Missouri 63103, USA}
\affiliation{Department of Physics, Michigan Technological University, Houghton, Michigan 49931, USA}

\author{Jan Wiersig}
\affiliation{Institut
    f{\"u}r Physik, Otto-von-Guericke-Universit{\"a}t Magdeburg, Postfach 4120,
    D-39016 Magdeburg, Germany} 
\date{\today}

\author{Ramy El-Ganainy}
\email{relganainy@slu.edu}
\affiliation{Department of Electrical and Computer Engineering, Saint Louis University, Saint Louis, Missouri 63103, USA}

\begin{abstract}
 \noindent
 One of the most fundamental questions in laser physics is the following: Which mode of an optical cavity will reach the lasing threshold first when gain is applied? Intuitively, the answer appears straightforward: When a particular mode is both temporally well confined (i.e., exhibits the highest quality factor) and experiences initially the largest increase of the modal gain, it is naturally expected to lase first. However, in this work, we demonstrate that this intuition can fail in surprising ways. Specifically, we show that in the presence of non-Hermitian degeneracies, known as exceptional points, the expected mode hierarchy can be dramatically altered. These spectral singularities can give rise to counterintuitive mode switching, where a mode with a lower quality factor and initially smaller increase of modal gain reaches the lasing threshold ahead of a more favorable competitor. Remarkably, this effect can occur even under spatially uniform pumping, underscoring the subtle and profound influence of non-Hermitian physics on lasing dynamics.
\end{abstract}

\maketitle

Laser systems of all types and scales have become indispensable tools across a broad spectrum of modern technologies. Beyond their wide-ranging practical applications, lasers also serve as powerful platforms for investigating fundamental phenomena in nonlinear dynamics \cite{SCM13,SS15}, wave localization \cite{Stano2013NPho,Liu2014NN}, and wave chaos \cite{Stone2001PS,CW15}, to just mention a few examples. Recent laser architectures have further enabled new regimes of mode selection~\cite{HodMirHas2016,SekOlySen2023} and robustness against disorder \cite{St-Jean2017NPho,Zhao2018NC,Bahari2017S,Bandres2018S}. While laser designs vary in material and structure, their operation follows general principles: (1) the mode with the highest quality factor $Q$ lases first under uniform gain; (2) mode competition above threshold can lead to complex dynamics \cite{Tureci2008S}; (3) single-mode lasing in large-area systems can be achieved via tailored pumping or dissipation \cite{CerRedGe2016,ChoSehKim2018, Ganainy2015PRA,Hokmabadi2019S,Qiao2021S}.

Against this backdrop, it was surprising to discover that spatial engineering of gain and loss in laser systems not only influences mode selection but can also give rise to counterintuitive phenomena such as gain-induced suppression of lasing and the closely related concept of loss-induced lasing \cite{LGC12,Brandstetter2014NC,POR14,Brandstetter2014NC,Ganainy2014PRA}. More recently, it has become clear that even uniform gain, without any spatial modulation, can give rise to rich and unexpected behaviors \cite{RiePanMak2021, HasBusOzd2022,KomMakBus2025}. 

\begin{figure}[t]
	\begin{center}
		\includegraphics[width=0.8 \columnwidth]{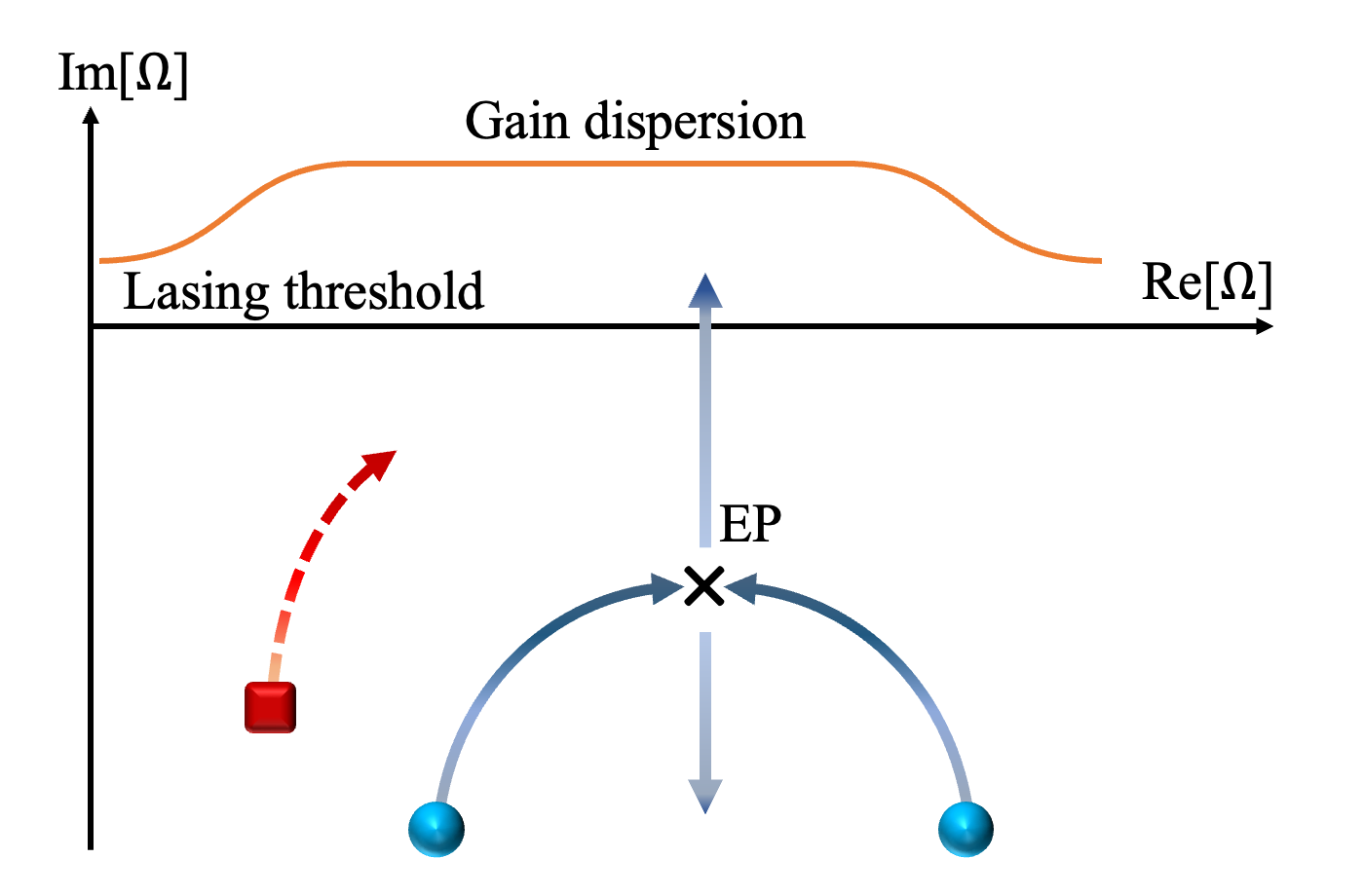}%
		\caption{A schematic illustration of the main concept of this work shows the distribution of complex eigenfrequencies~$\Omega$ of a laser cavity below the lasing threshold. The solid square and circles indicate the initial positions of three different eigenfrequencies in the absence of gain. The curved trajectories depict how these eigenfrequencies evolve as gain is introduced. Initially, the red mode (shown by the solid red square) has the lowest losses (i.e., the highest quality factor) and moves most rapidly (see dashed red line) toward the real axis as gain increases. However, at a certain gain level, two lower-quality modes (shown in solid blue lines) coalesce at an exceptional point (EP). Beyond this point, one of the resulting hybrid modes advances more quickly toward the real axis, eventually becoming the first to reach the lasing threshold—despite originating from lower-$Q$ modes. Ideally, the modes involved in this process should lie under the maximum value of a gain curve that excludes other modes but as we will see, this condition is not a necessary one.}
		\label{Fig_Schematic}
	\end{center}
\end{figure}

\begin{figure*}[!t]
\begin{center}
\includegraphics[width=\linewidth]{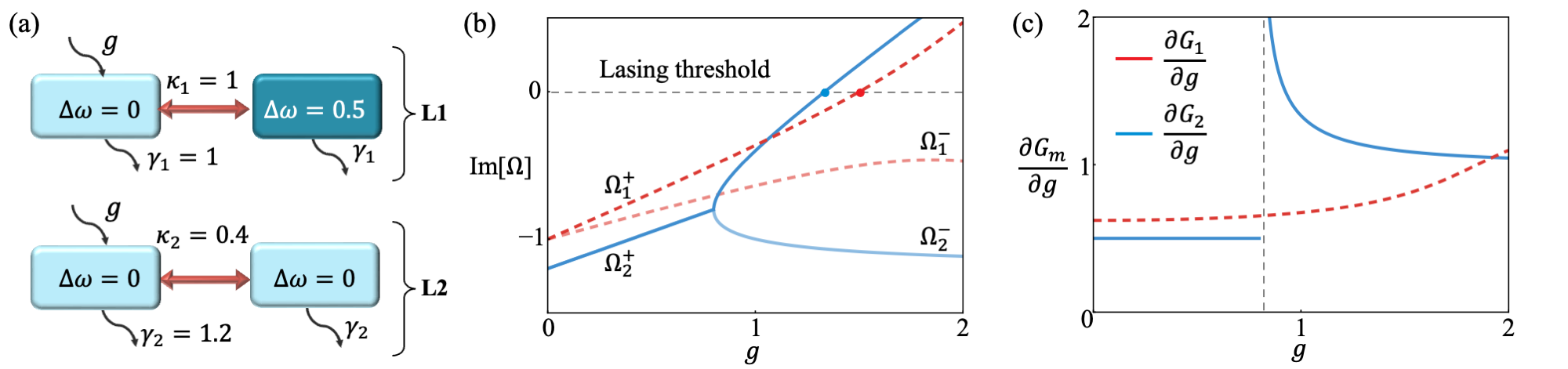}%
\caption{A simplified model illustrating the key effect discussed in this work can be constructed using two distinct laser systems, each consisting of two coupled cavities, as shown in (a). In both systems, the left cavity serves as the active laser element, while the right cavity acts as a passive reservoir. (b) The lasing mode of the first system (dashed red lines) has initially, i.e., in the absence of gain, a higher quality factor (lower loss $|\text{Im}\Omega|$)  than any of the modes in the second system. However, as the gain $g$ increases, the modes of the second system (solid blue lines) coalesce at an EP. Beyond this point, one of the resulting hybrid modes accelerates toward the lasing threshold and surpasses the initially higher-$Q$ mode of the first system. (c) The rate of modal gain increases as a function of applied gain shows a clear phase transition experienced by the blue mode, ultimately allowing it to lase before the blue mode. Although this model involves two separate laser systems, it provides an intuitive demonstration of the underlying mechanism. In the main Letter and Supplemental Material \cite{SM}, we show how this effect can be realized within a single laser system.}
\label{Fig_Schematic_CMT}
\end{center}
\end{figure*}

Despite the significant progress and the wealth of intriguing results related to non-Hermitian and topological effects in various laser systems,  to date, the golden rule of lasing mode selection still holds: under uniform pumping  and relatively dispersionless gain conditions within the frequency rage of interest, the mode with the highest quality factor and the largest modal gain change rate (a condition often overlooked) will be the first to reach the lasing threshold as the pump power is increased. This principle has been one of the crucial cornerstones of laser physics, providing a reliable guideline for predicting and engineering lasing mode behavior in a wide range of systems. In this study, we demonstrate a counterintuitive phenomenon that challenges this conventional expectation. Specifically, we demonstrate that in a uniformly pumped laser system, the optical mode with the highest quality factor and the initially largest modal gain change rate does not always reach the lasing threshold first. Contrary to this universal assumption, we show that in certain systems, as the uniform pump intensity is gradually increased from zero, two modes with lower quality factors may coalesce and form an exceptional point (EP)\cite{Kato66,MA19,Wiersig22}. Beyond this point, one of these modes exhibits a stronger increase of the modal gain and ultimately becomes the first to reach the lasing threshold. This behavior, which is shown schematically in Fig. \ref{Fig_Schematic}, underscores the intricate dynamics of mode competition driven by linear non-Hermitian effects taking place below the lasing threshold. In what follows, we discuss this effect using a toy model and full-wave simulations.\\

\noindent
\textit{Toy model}---We begin by considering a scenario that provides insight into the key results of this paper. Specifically, let us consider two distinct copies of a dimer laser system as shown in Fig. \ref{Fig_Schematic_CMT}(a), referred to as L1 and L2. In both cases, the active cavities share the same resonant frequencies and receive identical gain values. However, the systems differ in terms of their losses, coupling coefficients, and the resonant frequency of the passive cavity. Within the framework of temporal coupled mode theory, both systems can be described by a Hamiltonian matrix of the form
\begin{equation}
H_n=\begin{bmatrix}
    ig-i\gamma_n  & \kappa_n\\
    \kappa_n  & \Delta \omega_n-i\gamma_n
\end{bmatrix},\,\text{with }n=1,2
\end{equation}
which corresponds to L1 and L2, respectively. Here, $\gamma$ and $g$ represent the loss and material gain (or pump) while $\kappa$ and $\Delta \omega$ are the coupling strength and the detuning between the cavities. In the above, we have implicitly assumed that the gain remains approximately constant over the frequency range of interest, meaning that frequency-dependent effects of the gain medium do not play a significant role in the observed behavior. The eigenvalues of these matrices, denoted as $\Omega_1^{\pm}$ and $\Omega_2^{\pm}$ are the eigenfrequencies of the laser systems and given by the expression: $\Omega_n^{\pm}=\Delta \omega_n/2+i(g/2-\gamma_n)\pm \sqrt{\kappa_n^2+(\Delta \omega_n-ig)^2/4}$. 

We pose the following question: Under the pump specified by the above Hamiltonian, which of these lasers will reach the lasing threshold first? Before we answer this question, we note that the notion of ``uniform pumping'' is somewhat imprecise in this context, as the pump is applied only to part of the system. Nevertheless, this example serves to illustrate the main result of this Letter before we introduce a concrete system with truly uniform pumping. Let us now examine the system's behavior using the (dimensionless) parameters shown in the figure. The first lasing mode in either system corresponds to the mode with the smallest loss, i.e., the mode with the smallest $|\text{Im}[\Omega]|$ which in this case is $\Omega^{+}$. Thus, we focus on these modes. At $g=0$, we have $\text{Im}[\Omega_1^{+}]=-i\gamma_1=-i$ while $\text{Im}[\Omega_2^{+}]=-i\gamma_2=-1.2i$, indicating that the mode in L1 has a higher quality factor than that in L2 ($Q_1>Q_2$ where $Q_{1,2}$ are the corresponding quality factors of the modes, since the ratio of the real parts of eigenfrequencies is nearly unity). 
Furthermore, for $g/2<\kappa_2$, the modal gain increase with respect to the pump satisfies $\frac{\partial G_1}{\partial g}>\frac{\partial G_2}{\partial g}$, where $G_m\equiv\text{Im}[\Omega_m^{+}(g)]-\text{Im}[\Omega_m^{+}(g=0)]$. This follows from the fact that  $G_1=g/2 +\left| \text{Im}\left[\sqrt{\kappa_1^2+(\Delta \omega_1-ig)^2/4}\right]\right|$, while $G_2$ is given by $G_2= g/2$. In other words, the lasing mode in L1 has both a higher quality factor and a larger modal gain increase rate as a function of the pump compared to the lasing mode in L2. Yet, as illustrated in Fig. \ref{Fig_Schematic_CMT}(b), the lasing mode of L2 reaches the lasing threshold first (i.e., crosses the zero imaginary axis) as the pump $g$ is increased. This counterintuitive behavior is a direct consequence of the formation of an EP between the $\Omega_2^{\pm}$ modes, beyond which the model gain increase rate reverses its trend. Beyond the EP, the relationship between modal gain increase rates of the two modes undergoes a reversal, leading to the condition $\frac{\partial G_1}{\partial g}<\frac{\partial G_2}{\partial g}$ as shown in Fig. \ref{Fig_Schematic_CMT}(c) which ultimately boosts the mode  $\Omega_2^{+}$ to reach threshold first. 

Having demonstrated the core concept, we proceed to demonstrate that the effect arises even under strictly uniform pumping conditions.\\

\noindent
\textit{EP-induced lasing mode switching in a single microcavity with uniform pumping}---Consider a chip-scale laser consisting of a single dielectric cavity. If the cavity is sufficiently thin in the vertical direction, the problem can be effectively treated as two-dimensional~\cite{Jackson83eng,CW15}. In the absence of pumping or gain—i.e., under passive conditions—the cavity modes are determined by solving the eigenvalue equation
\begin{equation}
	\label{eq:modeeq}
	\left[\Delta_{\perp} + n^2(x,y)k^2\right] \psi = 0
\end{equation}
where $\Delta_{\perp}$ is the transverse Laplacian, $\psi$ represents either the out-of-plane component of the electric or  the magnetic field as either $\vec{E}$ or $\vec{H}$ are given by $(0, 0,\text{Re}[\psi(\vec{r})e^{-i\omega t}])^{\textsf{T}}$; $\omega=c k$ is the eigenfrequency and $c$ is the speed of light in vacuum. The refractive index distribution, $n(x,y)$ defines the cavity and extends over a finite, connected region, outside of which the refractive index is unity. The cavity geometry can take various forms, such as a microdisk \cite{MLSGPL92,CaoXuXia2000,LiuFanHua2004}, Lima\c{c}on-shaped \cite{WieHen2008,SCL09, WYD09,ShiHenWie2009,YiKimKim2009,AHE12}, spiral \cite{CheTurSto2003,BenZys2005}, or a stadium-shaped cavity \cite{FukHar2004,FanYamCao2005}. 

To determine the modes, Eq.~\eqref{eq:modeeq} is solved under outgoing boundary conditions~\cite{CW15}. This open boundary nature renders the problem non-Hermitian, leading to eigenmodes that are not true bound states but rather quasinormal modes (QNMs) \cite{CLM98} with complex eigenfrequencies $\omega$. The negative imaginary part of $\omega$ determines the temporal decay rate of the mode. For convenience, the eigenfrequency $\omega$ is typically scaled by a characteristic length $R$ of the system, yielding the dimensionless quantity $\Omega = \omega R/c = kR$. In terms of this scaled frequency, the quality factor of a given mode is defined as $Q=-\text{Re}[\Omega]/(2\text{Im}[\Omega])$. 

When gain is introduced, the cavity transitions from a passive to an active system. This effect is incorporated by assigning a negative imaginary component to the refractive index $n(x,y)$ within the cavity region, while the surrounding medium remains free space or a uniform passive substrate. The new eigenfrequencies $\Omega$ are then determined by solving the modified eigenvalue problem. The lasing threshold for a given QNM is reached when gain exactly compensates for loss, corresponding to $\text{Im}[\Omega]=0$ or equivalently $Q=\infty$. \\

With the problem now defined, let us examine how the application of gain gradually affects a passive dielectric cavity. In general, two key effects emerge: (1) an increase in the quality factors of the modes and (2) a modification of the electromagnetic field distributions associated with each mode. While these effects are inherently coupled, their relative dominance depends on the cavity geometry and operating conditions.

\begin{figure}[tb]
	\begin{center}
		\includegraphics[width= \columnwidth]{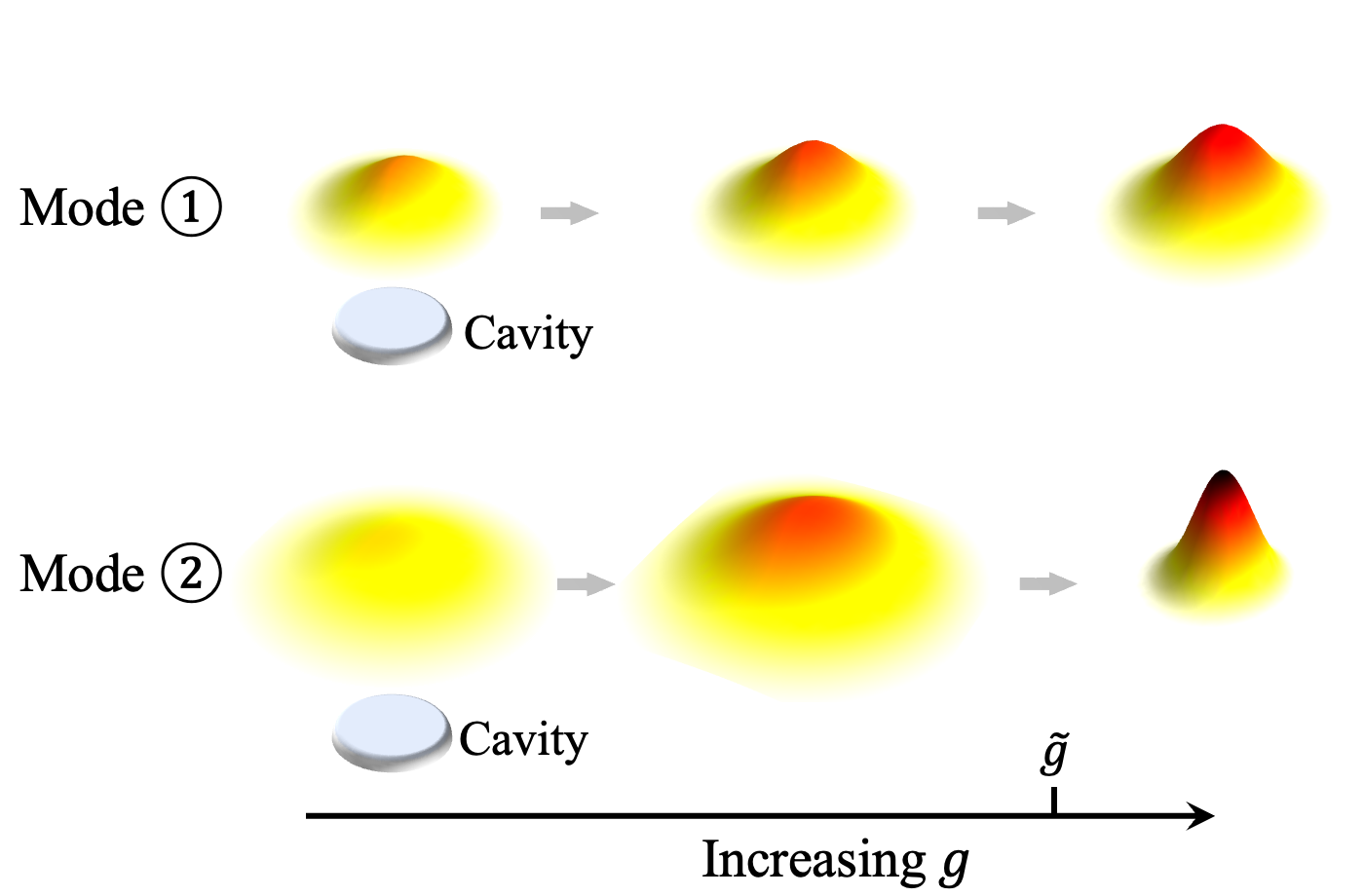}%
		\caption{A schematic illustration of the concept discussed in this work is shown in the context of a single laser cavity under uniform pumping. In the passive regime, mode \textcircled{1} is initially more confined within the cavity volume, resulting in a higher quality factor compared to mode \textcircled{2}. As gain~$g$ is applied, mode \textcircled{1} initially experiences a stronger increase of modal gain and approaches the lasing threshold more quickly. However, if mode \textcircled{2} coalesces with another quasinormal mode (not shown here) to form an EP at a certain gain threshold $\tilde{g}$, the resulting hybrid mode can become more spatially confined than the mode \textcircled{1}. As a result, mode \textcircled{2} may experience a stronger increase of modal gain and ultimately reach the lasing threshold first. }
		\label{Fig_SC_Concept}
	\end{center}
\end{figure}

Figure \ref{Fig_SC_Concept} presents a schematic illustrating two distinct scenarios. In this representation, the disks correspond to dielectric cavities, and the Gaussian profile represents different modes. For mode \textcircled{1}, it is assumed that increasing the gain does not significantly alter its spatial profile or confinement factor. As a result, its modal gain increases gradually without abrupt structural changes. In contrast, mode \textcircled{2} follows a markedly different trajectory: above a certain gain threshold $\tilde{g}$, it undergoes a sudden structural change, leading to a sharp reduction in its confinement factor and a corresponding rapid increase in modal gain. Consequently, while mode \textcircled{1} may initially possess both a higher quality factor and a stronger increase in modal gain with applied gain, the phase transition associated with mode \textcircled{2} can ultimately cause it to reach the lasing threshold first. Such a phase transition is necessarily accompanied by the formation of an EP involving mode \textcircled{2} and another mode in the system (not shown here), further underscoring the critical role of non-Hermitian physics in governing lasing dynamics. We emphasize that this phenomenon occurs under uniform pumping conditions, distinguishing it from scenarios where the spatial profile of the applied gain is deliberately engineered to favor a lower-quality-factor mode over one with a higher quality factor. In our case, the mode selection arises intrinsically from the system’s dynamics rather than engineering the external pump.\\

\begin{figure}[tb]
	\begin{center}
		\includegraphics[width=0.89\columnwidth]{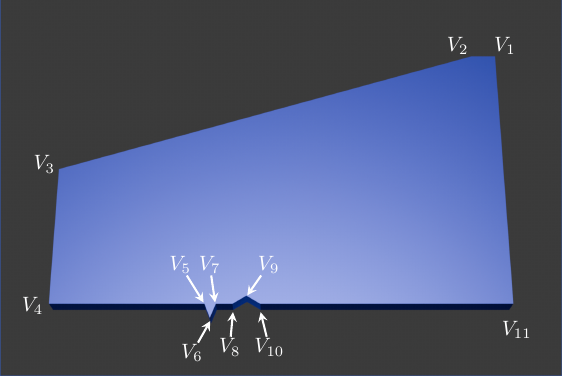}%
		\caption{An illustration of the proposed microcavity. The polygonal shape is given by the vertices $V_{1-11}$ with relative coordinates $(x/R,y/R)$ being $(1, 0.59637)$, $(0.88924, 0.59637)$, $(-1, 0.02815)$, $(-1, -0.59637)$, $(-0.33022, -0.59637)$, $(-0.30598, -0.65828)$, $(-0.28174, -0.59637)$, $(-0.21017, -0.59637)$, $(-0.15, -5.6009)$, $(-0.08983, -0.59637)$, and $(1, -0.59637)$, respectively.}
		\label{fig:CavityShape}
	\end{center}
\end{figure}

To validate the occurrence of the intriguing scenario described above, we designed a polygonal microcavity with eleven vertices, denoted as $\text{V}_{1-11}=(x_{1-11}/R, y_{1-11}/R)$ as illustrated in Fig.~\ref{fig:CavityShape}. The exact coordinates of these vertices are provided in the figure caption. The cavity is characterized by a complex refractive index of $n=2.1+i n_{\text{imag}}$ where a negative value of $n_{\text{imag}}$ corresponds to optical gain, as previously described (we neglect gain dispersion here to emphasize and isolate the non-Hermitian effect and demonstrate its robustness across the frequency range of interest). The surrounding medium is assumed to be free space with a refractive index of  $n=1$. Therefore, the refractive index of the cavity represents commonly used microcavity laser materials like zinc oxide~\cite{NKRLG04}, lithium niobate~\cite{LYW2020,LYZ2021,GGY2021} or aluminum nitride~\cite{LSX2017,LBG2018}. 

\begin{figure}[tb]
	\begin{center}
		\includegraphics[width=\columnwidth]{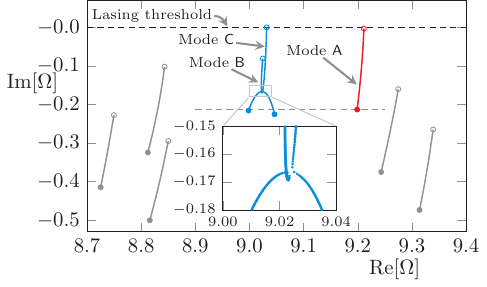}%
		\caption{Complex frequency plane. The frequency trajectories of modes in a polygonal cavity are shown for a change of the imaginary part of the refractive index from $n_{\text{imag}}=0$ (filled circles) to $n_{\text{imag}}=-0.04977$ (empty circles). The inset is a magnification with results of the individual simulations shown by thin dots.}
		\label{fig:ComplexFreq}
	\end{center}
\end{figure}

Figure~\ref{fig:ComplexFreq} shows the trajectories of various complex eigenfrequencies of the system as a function of gain, i.e., $n_{\text{imag}}$. These solutions were obtained by numerically solving Eq.~(\ref{eq:modeeq}) for QNM $\psi$ representing a magnetic field perpendicular to the cavity plane using the finite-element solver JCMwave \cite{PomBurZsc2007}, focusing on modes within the frequency range centered around $\text{Re}[\Omega] \sim 9.0$. As shown in the figure, when $n_{\text{imag}}=0$, a mode labeled {\small\textsf{A}}, with $\text{Re}[\Omega] \approx 9.2$, exhibits the highest $Q$ factor as it is closest to the real axis. Additionally, two other relevant modes, labeled {\small\textsf{B}} and {\small\textsf{C}}, within the range $8.9 \leq \text{Re}[\Omega] \leq 9.1$, possess slightly lower quality factors.

\begin{figure}[tb]
	\begin{center}
		\includegraphics[width=\columnwidth]{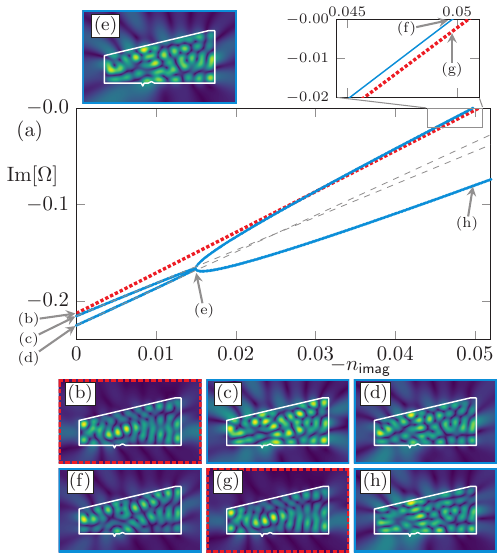}%
		\caption{Gain dynamics of optical modes. (a) Imaginary part of $\Omega$ versus the gain implemented as negative imaginary part of the refractive index for long-lived modes with $\text{Re}[\Omega]\approx9.0$. A red dotted curve represents the mode which initially has the largest quality factor and largest increase in modal gain. Blue solid curves are shown for a mode pair that coalesces into an EP at $n_{\text{imag}} = -0.015$. The gray dashed curves are the estimated behavior of $\text{Im}[\Omega]$ based on the modal gain increase rate at $n_{\text{imag}}=0$. Mode patterns $|\psi|$ are shown in (b)--(d) for $n_{\text{imag}}=0$, (e) for $n_{\text{imag}}=-0.015$, and (f)-(h) for $n_{\text{imag}}=-0.04977$. }
		\label{fig:ImagFreq}
	\end{center}
\end{figure}

As the gain in the cavity gradually increases, all modes shift in the complex frequency plane toward the real axis. The modal gain increase $\frac{\partial G_m}{\partial g}$ with $G_m=\text{Im}[\Omega]$ and $g=-n_{\text{imag}}$ quantifies how rapidly each mode approaches the real axis and can be evaluated at each gain value. For the passive cavity with $n_{\text{imag}}=0$, we find $\frac{\partial G_\text{A}}{\partial g}=4.19$ while $\frac{\partial G_\text{B,C}}{\partial g}<3.81$. Based on this initial estimation, which considers only the quality factor and the initial increase of modal gain, one would expect mode {\small\textsf{A}} to reach the lasing threshold first, followed by modes {\small\textsf{B}} and {\small\textsf{C}}. However, as indicated by dashed lines in Fig.~\ref{fig:ImagFreq}(a), this prediction turns out to be incorrect. Instead, at $n_{\text{imag}}=-0.015$ the modes {\small\textsf{B}} and {\small\textsf{C}} coalesce at an EP; see the inset of Fig.~\ref{fig:ComplexFreq}. Beyond this point, the modal gain increase of mode~{\small\textsf{C}} exceeds that of mode {\small\textsf{A}}, i.e., $\frac{\partial G_\text{C}}{\partial g}>\frac{\partial G_\text{A}}{\partial g}$ [see blue curves in Fig.~\ref{fig:ImagFreq} (a)]. Consequently, the modal gain of mode {\small\textsf{C}} surpasses that of mode {\small\textsf{A}}, allowing it to reach the lasing threshold first at $n_{\text{imag}}=-0.04977$. As discussed earlier, this behavior is related to the mode confinement factor, which we examine in more detail in the Supplemental Material \cite{SM}. Figures \ref{fig:ImagFreq}(b)--(f) also present the near-field field distributions $|\psi|$ associated with modes {\small\textsf{A}} and {\small\textsf{C}} for different gain values.  Finally, the inset in Fig.~\ref{fig:ImagFreq}(a) showing a magnification around the lasing threshold, evidently indicates the switching of modes {\small\textsf{A}} and {\small\textsf{C}}. 
For completeness, we have also investigated the robustness of the lasing mode switching effect under deviations from the sharp polygonal shape of the cavity. As detailed in the Supplemental Material \cite{SM}, our results indicate that the effect is stable up to a corner rounding with a radius $r_{\text{rounding}}\approx0.01R$. In the Supplemental Material \cite{SM}, we additionally discuss the evolution of the far-field emission pattern when the gain is increased. 

Finally, while the structure studied in Fig.~\ref{fig:CavityShape} is experimentally realizable, it does not represent a standard laser geometry in integrated photonics. We present an alternative configuration using a more conventional platform in the Supplemental Material \cite{SM}.\\


In conclusion, we have demonstrated that the intuitive expectation—that the first lasing mode corresponds to the highest-quality mode with the largest initial gain increase—is not always valid. Exceptional points below threshold can induce mode switching, allowing a lower-quality mode to lase first under uniform pumping. This effect, illustrated by a coupled waveguide model and validated by full-wave microcavity simulations, highlights the crucial role of non-Hermitian physics in lasing dynamics and mode control. While the effect presented in this work does not necessarily offer a practical method for laser mode selection, it highlights the rich physics of non-Hermitian systems and emphasizes the importance of accounting for these effects in laser design to avoid misleading or faulty predictions.\\

\begin{acknowledgments}
\textit{Acknowledgments}--R.E.  acknowledges support from the AFOSR Multidisciplinary University Research Initiative Award on Programmable Systems with Non-Hermitian Quantum Dynamics (Grant No.FA9550-21-1-0202) and from the Army Research Office (W911NF-23-1-0312). R.E. and J.K. gratefully acknowledge the Mathematisches Forschungsinstitut Oberwolfach (MFO) for hosting the workshop \textit{Nonlinear Optics: Physics, Analysis, and Numerics}, where the discussions that led to this work first began.
\end{acknowledgments}

%

\clearpage

\section{Supplementary Materials}
\renewcommand{\thefigure}{S\arabic{figure}}
\renewcommand{\theequation}{S\arabic{equation}}
\setcounter{figure}{0}
\setcounter{equation}{0}

\subsection{1. Power confinement factor}
\begin{figure}[!htb]
	\begin{center}
		\includegraphics[width=\columnwidth]{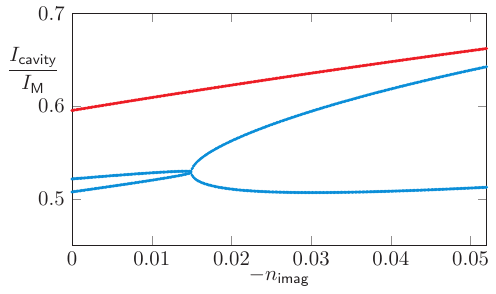}
		\caption{Confinement ratio of integrated intensity inside the cavity and overall intensity in the computational domain for (red curve) mode {\textsf{A}} and (blue curves) modes {\textsf{B}} and~{\textsf{C}}.}
		\label{fig:SMConfinementRatio}
	\end{center}
\end{figure}

Figure 3 in the main text schematically illustrates the effect of laser mode switching under applied gain $g=-n_{\text{imag}}$ in terms of mode confinement in the cavity. The sketched behavior can be observed in the polygonal cavity as well. To do so, we consider a finite computational domain $\textsf{M}$ around the cavity. With respect to $\textsf{M}$ the quasi normal mode can be normalized such that the confinement ration 
\begin{equation}
	\frac{I_{\textsf{cavity}}}{I_{\textsf{M}}} = \frac{\int_{\textsf{cavity}}\,|\psi(\vec{r})|^2\text{d}^2r \,}{\int_{\textsf{M}}\, |\psi(\vec{r})|^2\,\text{d}^2r }
\end{equation}
is finite. Note that quasi-normal modes typically cannot be normalized with respect to the full $\mathbb{R}^2$, as they diverge at an infinite distance from the cavity. Thus, mathematically, the confinement ratio is zero for $\textsf{M}=\mathbb{R}^2$. However, we choose a more practical approach by considering $\textsf{M}=\lbrace (x, y) | -2\leq x/R \leq 2 \text{ and } -1\leq y/R \leq 1 \rbrace$, where again $R$ is a characteristic length of the system.
As shown in Fig.~\ref{fig:SMConfinementRatio} mode {\textsf{A}} and modes {\textsf{B}} and~{\textsf{C}} have a fundamentally different behavior under applied gain: Mode {\textsf{A}} has a linear dependence of the confinement ratio with the applied gain while  modes {\textsf{B}} and~{\textsf{C}} are initially almost equally confined but split after the coalescence at the EP into a strongly confined and a weakly confined mode. Thus, the behavior of the confinement ratio for modes {\textsf{B}} and~{\textsf{C}} is nonlinear and resembles the square-root topology corresponding to the modal gain shown in Fig.~6 in the main text.

Note that the qualitative values of the confinement ratio depend on the shape and size of the computational domain $\textsf{M}$ and the particular mode pattern in this finite region. Thus, a quantitative comparison between mode {\textsf{A}} and modes {\textsf{B}} and~{\textsf{C}} differ by changing  $\textsf{M}$ but the qualitative behavior of the confinement factor remains.

\subsection{2. Robustness of lasing mode switching against corner rounding}
In Fig.~6 of the main text, the effect of the lasing mode switching is evidently shown for a microcavity with a polygonal shape defined by the vertices $V_{1-11}$. The shape of that cavity can also be seen in Fig.~4 of the main text. Its design does not follow a specific algorithm but is rather a combination of intuition and numerical optimization. Especially, for a numerical fine tuning  of an individual mode pair the small notches are advantageous as they provide enough parameters to achieve an EP with correctly oriented branch cut curves in parameter space. At the same time, the overall shape of the cavity is chaotic such that the modes distribute in the whole cavity which gives more rich possibilities to reshuffle intensity for the laser mode switching effect.

\begin{figure*}[tb]
\begin{center}
\includegraphics[width=0.99\textwidth]{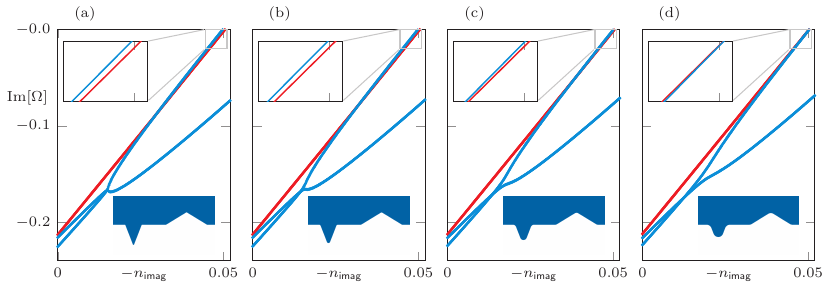}%
\caption{Robustness against corner rounding of the polygon. In comparison to Fig.~5 in the main text, the behavior of $\text{Im}\,\Omega$ with increasing gain is shown for cavities with different corner rounding of the polygonal shape. In (a) to (d) corners are rounded with a radius $0.002R$, $0.005R$, $0.01R$, and $0.015R$, respectively. The inset in the lower left of  each diagram shows a magnification of the microcavity around the double-notch defined by vertices $V_{5-10}$ (cf.\ Fig.~4 in the main text).}
\label{fig:SMCornerRounding}
\end{center}
\end{figure*}

However, a polygonal shape with sharp corners is a rather idealized scenario. In a realistic setup, corners have to be rounded to resemble the fabrication tolerances. In Fig.~\ref{fig:SMCornerRounding} the impact of the corner rounding on the lasing mode switching effect is shown for varying to rounding radii from $r_{\text{rounding}}=0.002R$ to $r_{\text{rounding}}=0.015R$. As can be seen the effect of lasing mode switching persists clearly up to $r_{\text{rounding}}=0.01R$ in Fig.~\ref{fig:SMCornerRounding} (c) where the corner rounding is easily noticeable in the geometry by eye. However, the corner rounding leads to a successive deviation from the EP at $n_{\text{imag}}=-0.015$ and therefore diminishing its influence on the non-linear behavior of the complex frequency. As a consequences, for a large rounding radii as of $r_{\text{rounding}}=0.015R$ in Fig.~\ref{fig:SMCornerRounding} (d) the lasing mode switching effect is dissolved. However, it should be mentioned that this effect can be compensated by re-optimizing the vertices $V_{5-10}$ under the assumption of a particular corner rounding.

\begin{figure}[b]
	\begin{center}
 		\includegraphics[width=\columnwidth]{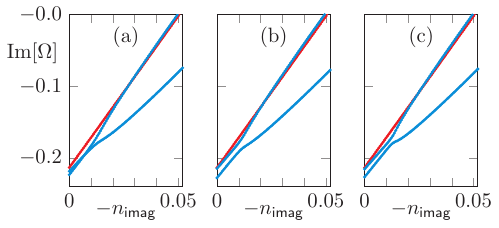}%
 		\caption{Gain dynamics of optical modes in a perturbed polygonal cavity. (a) The left notch is moved $\Delta x/R = 0.01$ to the right. (b) The base width of the left notch is increased by $\Delta w/R = 0.01$. (c) The height the left notch is increased by $\Delta h/R = 0.005$.}
 		\label{fig:SMNotch1Pert}
	\end{center}
\end{figure}

Note that corner rounding is a natural and generic perturbation to the cavity shape. Therefore, we expect the laser mode switching to be robust against other parameter variations as well. For verification we changed the position, width and height of the notch defined by vertices $V_{5-7}$ (left notch). As shown in Fig.~\ref{fig:SMNotch1Pert}, the laser mode switching is preserved even though the EP-degeneracy is lifted.

The behavior of the complex frequencies in a perturbed polygonal cavity, e.g. via corner rounding or notch displacement, is well described within the toy model of the main text. By considering a complex $\Delta \omega_2$ in the the Hamiltonian matrix (1) the EP degeneracy is lifted and the curves for $\Omega_2^\pm$ as function of the applied gain become smooth. As long as $|\Delta \omega_2|$ is small enough, the laser mode switching effect is still present.

\subsection{3. Far-field radiation of the polygonal cavity}
\begin{figure*}[tb]
\begin{center}
\includegraphics[width=0.92\textwidth]{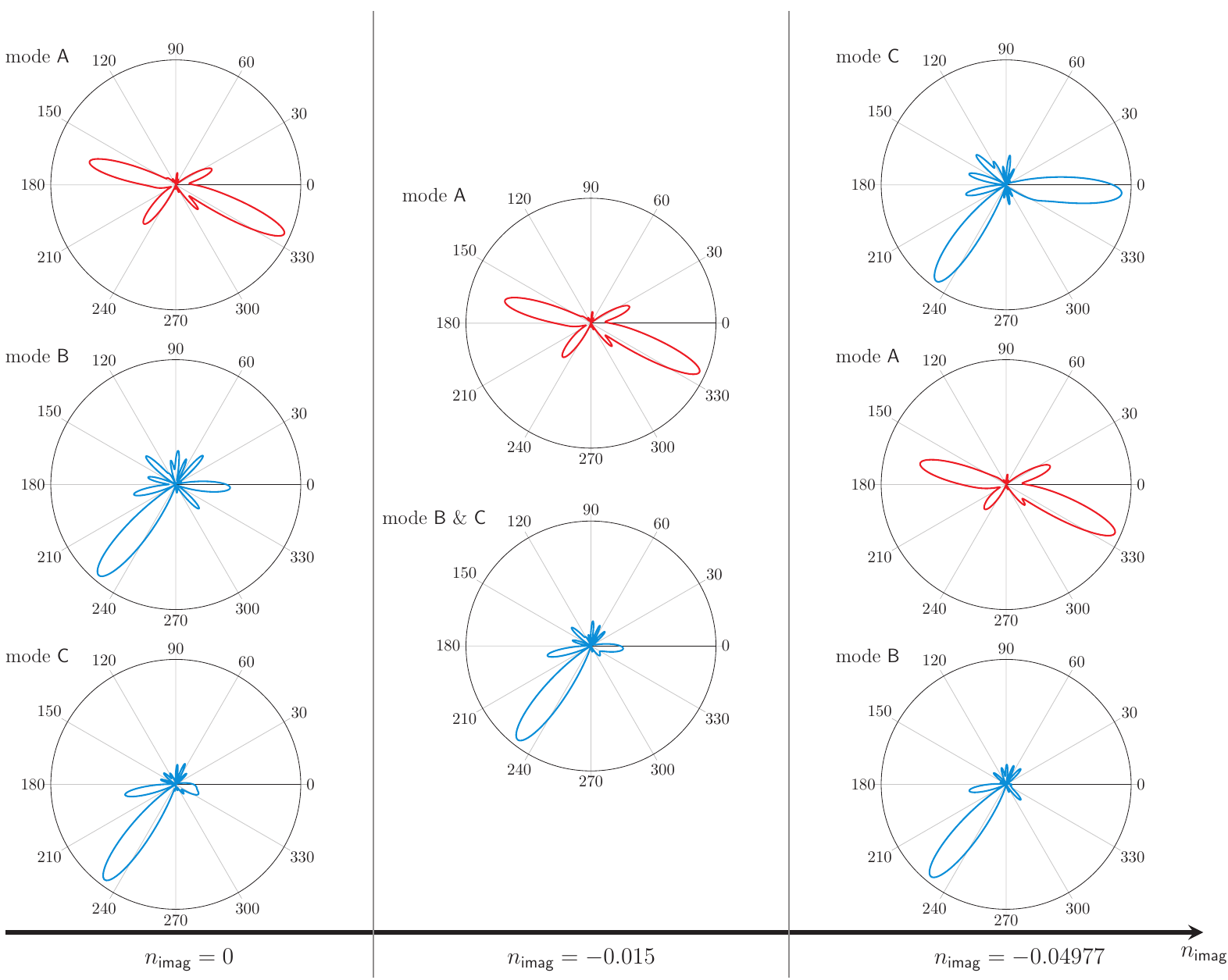}%
\caption{Electric farfield intensity pattern of modes \textsf{A}, \textsf{B}, and \textsf{C} in the polygonal cavity for different values of $n_{\text{imag}}$. Each pattern is normalized to the maximum value.}
\label{fig:SMFarfield}
	\end{center}
\end{figure*}

In a potential experiment it is essential to distinguish light emission from different modes to verify the effect of laser mode switching. In this section we therefore show that the intensity of the electric farfield radiated in the cavity plane can be used to identify the lasing mode. As shown in Fig.~\ref{fig:SMFarfield} in a cavity without gain, i.e.\ with $n_{\text{imag}}=0$, the high-quality mode {\textsf{A}} radiates in diagonal directions from the cavity whereas modes {\textsf{B}} and~{\textsf{C}} predominantly emit in the direction of 230$^\circ$ with two minor emission peaks in 0$^\circ$ and 190$^\circ$ direction. When gain is applied to the cavity the radiation pattern of mode {\textsf{A}} remains almost unchanged. On the other hand, the modes {\textsf{B}} and~{\textsf{C}} show a pronounced change of their radiation patterns under the variation of the applied gain. First, these modes coalesce at the EP at $n_{\text{imag}}=-0.0015$ and then rearrange their fields to hybridize in a well-confined and a weakly confined mode. This rearrangement of the fields is accompanied with a change in the emission patterns: mode \textsf{C}  retains the minor emission peak at 0$^\circ$ and only contributes weakly to the 230$^\circ$ direction. In contrast, mode \textsf{B} dominates the emission in both the 230$^\circ$ and 190$^\circ$ directions. In particular, the well-marked radiation of mode {\textsf{C}} in the zero degree direction can be used for identification.

\clearpage

\subsection{4. Implementation of Two Coupled Lasers in a Composite Cavity System}

\begin{figure}[tb]
\begin{center}
\includegraphics[width=\columnwidth]{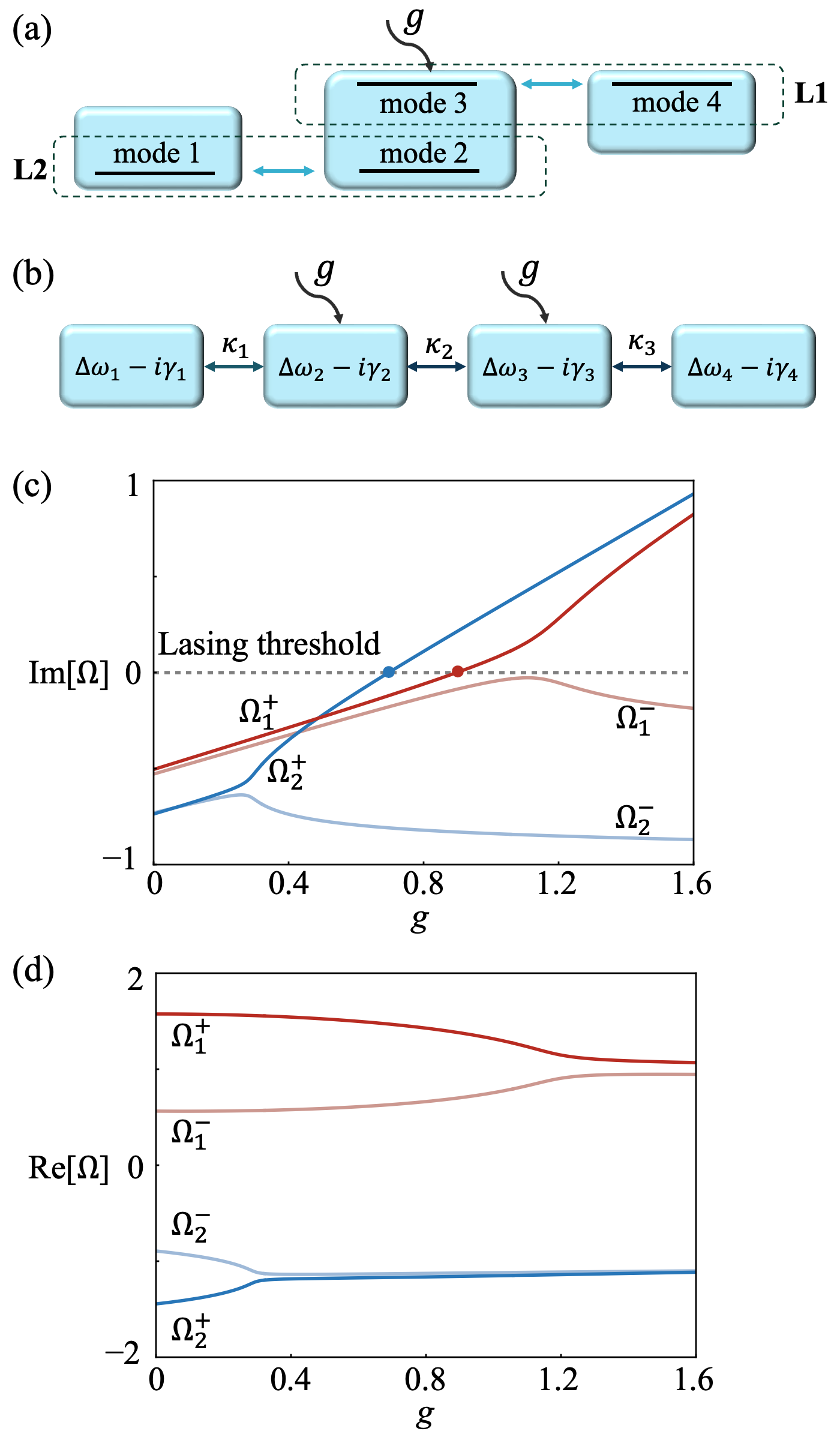}%
\caption{(a) Implementation of both lasers L1 and L2 (see Fig.~2 of the main text) within a single structure. The central cavity supports two spatial modes, labeled 2 and 3, which couple to mode 1 in the left cavity and mode 4 in the right cavity, respectively, enabling the formation of the two lasers. (b) One possible realization of the level structure in (a) using four coupled single-mode cavities. The parameters are chosen to match the full-wave simulation results shown in Fig.~\ref{Fig-SM-4cavities_Simulation}: $\Delta \omega_1 = -1.05$, $\Delta \omega_2 = \Delta \omega_3 = 0$, $\Delta \omega_4 = 0.85$; $\gamma_1 = 1$, $\gamma_2 = \gamma_3 = \gamma_4 = 0.5$; 
$\kappa_1 = 0.6$, $\kappa_2 = 1$, $\kappa_3 = 0.8$. (c) Imaginary parts of the eigenfrequencies of the four coupled-cavity supermodes vs gain~$g$. Red and blue dots indicate the lasing thresholds of the corresponding modes.
}
\label{Fig-SM-4cavities_CMT}
\end{center}
\end{figure}

\begin{figure}[t]
\begin{center}
\includegraphics[width=\columnwidth]{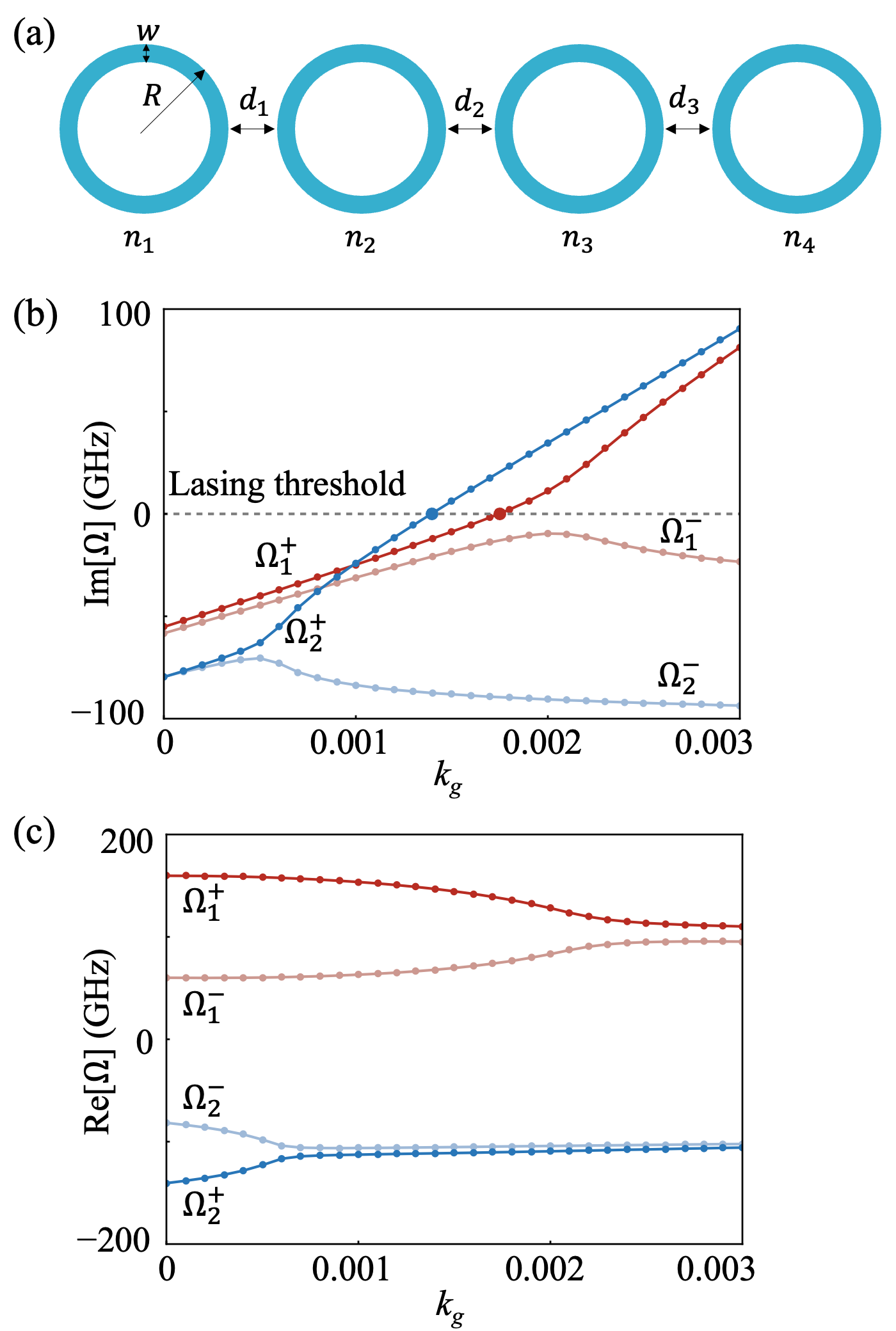}%
\caption{(a) Geometry of the four microring cavities used in the simulation. All rings have identical dimensions: radius $R = 5\,\mu\text{m}$ and width $w = 0.25\,\mu\text{m}$. The center-to-center distances between adjacent rings are $d_1 = 0.23\,\mu\text{m}$, $d_2 = 0.19\,\mu\text{m}$, and $d_3 = 0.21\,\mu\text{m}$.  The complex refractive indices of the rings core are defined as $n_j = n_\text{core} + \Delta n_j - i k_j$, where $n_\text{core} = 3.5$, $\Delta n_1 = 0.0018$, $\Delta n_4 = -0.00155$, and $\Delta n_2 = \Delta n_3 = 0$; the corresponding loss coefficients are $k_1 = 0.002$ and $k_2 = k_3 = k_4 = 0.001$. The surrounding medium  is assumed to be free space with $n_\text{clad} = 1$. Optical gain is introduced in the two central rings by adding an imaginary component $i k_g$ to $n_{2}$ and $n_{3}$. (b) Imaginary parts of the eigenfrequencies of the four coupled-cavity supermodes obtained from full-wave simulation.
}
\label{Fig-SM-4cavities_Simulation}
\end{center}
\end{figure}

\begin{figure*}[t]
\begin{center}
\includegraphics[width=6in]{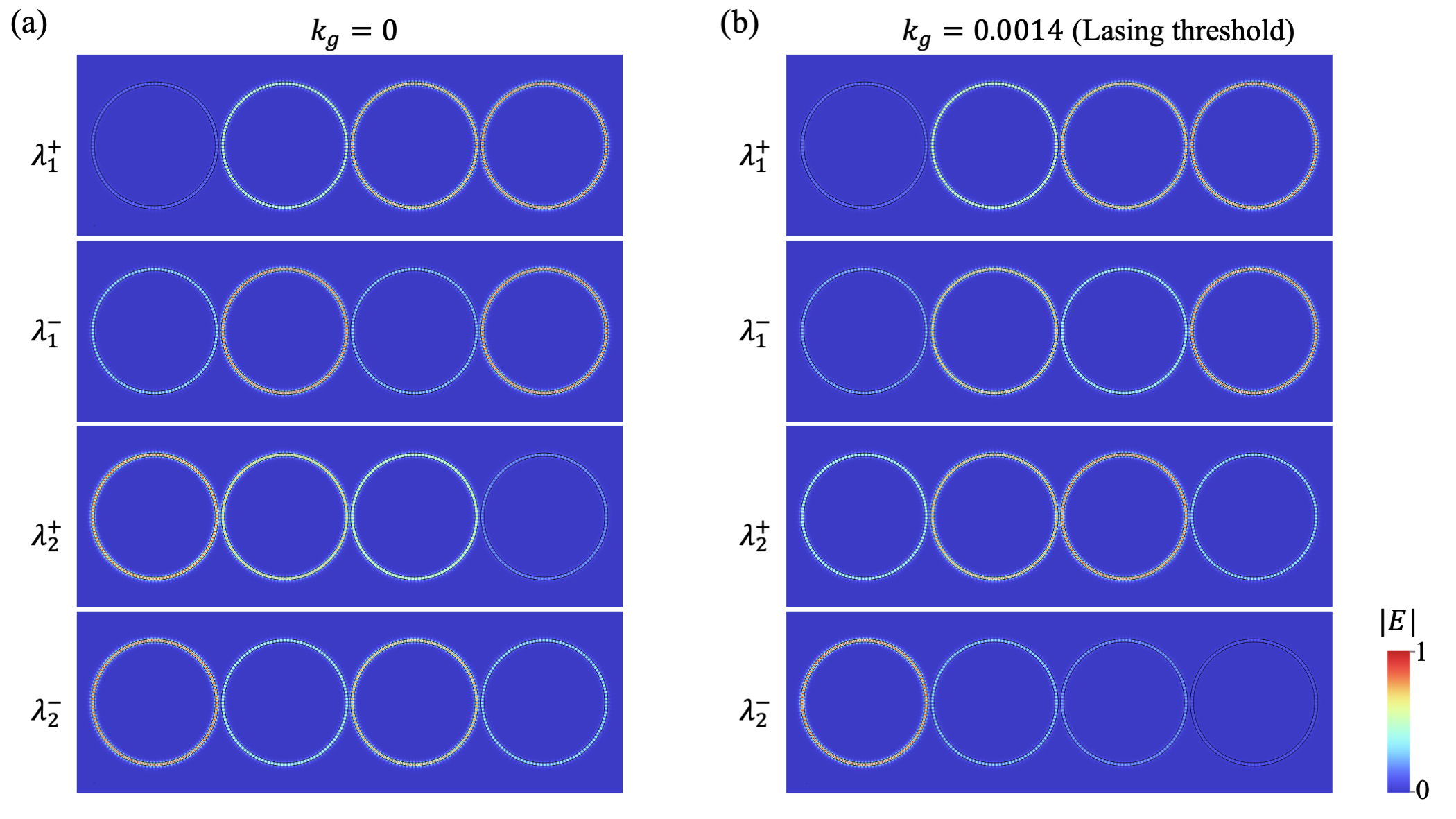}%
\caption{(a) Spatial profiles of the four supermodes in the coupled cavity system shown of Fig.~\ref{Fig-SM-4cavities_Simulation}(a), in the absence of gain. (b) Spatial profiles of the four supermodes when a gain of $k_g = 0.0014$ is applied to the two central microring cavities, corresponding to the lasing threshold of the $\Omega_2^+$ mode.
}
\label{Fig-SM-Modes}
	\end{center}
\end{figure*}

In Fig. 2 of the main text, lasers L1 and L2 are modeled as two independent systems, each consisting of a pair of coupled single-mode cavities. This separation streamlines both the theoretical model and its discussion. However, an equivalent configuration can be implemented using a single system of three coupled cavities, as illustrated in Fig. \ref{Fig-SM-4cavities_CMT}(a). In this setup, the central cavity supports two spatially overlapping modes (modes 2 and 3). Mode 3 couples to mode 4 in the right cavity to form laser L1, while mode 2 couples to mode 1 in the left cavity to form laser L2. In general, one can also include weak cross-couplings between mode 1 and mode 3, and between mode 4 and mode 2. Nonetheless, due to their large frequency detunings, these interactions have a negligible effect on the lasing dynamics. In this three-cavity implementation, modes 2 and 3 play the role of the left cavity mode in Fig. 2 of the main text, whereas modes 1 and 4 correspond to the right cavity mode.

While it is natural for a single cavity to support multiple modes, a more effective strategy for engineering a two-mode cavity is to use a pair of identical single-mode cavities coupled to each other. The mutual coupling results in hybridization, producing two supermodes whose frequency splitting depends on the coupling strength. This approach offers greater control over both the frequency separation and the spatial profiles of the resulting modes. Figure~\ref{Fig-SM-4cavities_CMT}(b) illustrates this implementation: four single-mode cavities, each characterized by a resonant frequency detuning $\Delta \omega_n$ and decay rate $\gamma_n$ ($n = 1, 2, 3, 4$), are coupled with strengths $\kappa_m$ ($m = 1, 2, 3$). Figure~\ref{Fig-SM-4cavities_CMT}(c) plots the imaginary parts (representing gain or loss) of the eigenvalues associated with the four eigenmodes of this system, based on the design parameters specified in the figure caption. At $g = 0$, the mode corresponding to $\Omega_2^+$ exhibits higher loss than the mode corresponding to $\Omega_1^+$. Nevertheless, due to `phase transition' taking place at the quasi-EP near $g = 0.3$, $\Omega_2^+$ reaches the lasing threshold first.

To confirm that the proposed concept can be realized in practical laser systems, we implemented the design using microring resonators and performed full-wave 2D simulations with realistic material parameters corresponding to the AlGaAs material platform. The geometry and refractive indices of the cavities are provided in the caption of Fig.~\ref{Fig-SM-4cavities_Simulation}. The frequency detuning and loss rates in each cavity are modeled by assigning a complex refractive index of the form $n_j = n_\text{core} + \Delta n_j + i k_j$ ($j = 1, 2, 3, 4$). Optical gain is introduced in the two central rings by adding an imaginary component $-i k_g$ to $n_{2,3}$ . The imaginary parts of the eigenfrequencies obtained from the simulation are plotted in Fig.~\ref{Fig-SM-4cavities_Simulation}(b), demonstrating good agreement with the theoretical predictions shown in Fig.~\ref{Fig-SM-4cavities_CMT}(c).

Figures~\ref{Fig-SM-Modes}(a) and (b) plot the spatial profiles of the supermodes at $k_g = 0$ and $k_g = 0.0014$ (the lasing threshold of L2), respectively. The $\Omega_1^{\pm}$ modes remain largely unchanged and are predominantly localized in the right three microrings. In contrast, the $\Omega_2^{\pm}$ modes are initially confined to the left three microrings. As gain is introduced, the $\Omega_2^+$ mode becomes increasingly concentrated in the two central rings where the gain is applied which explains why it reaches the lasing threshold first. Meanwhile, the $\Omega_2^-$ mode shifts toward the leftmost ring, which has the highest loss, consistent with its large decay rate shown in Fig.~\ref{Fig-SM-4cavities_Simulation}(b).

\end{document}